\documentstyle[preprint,aps]{revtex}
\begin{document}

\title{
Coulomb Drag Between Parallel\\ Ballistic Quantum Wires}

\author{  O. Raichev\cite{oleg} and P. Vasilopoulos\cite{takis}\\
\ \\}

\address{\cite{oleg}Institute of Semiconductor Physics,
National Academy of Sciences,\\ 45 Prospekt Nauki, Kiev 252650,  Ukraine\\
\ \\
\cite{takis}Concordia University, Department of Physics,
1455 de Maisonneuve Ouest,\\ Montr\'{e}al, Qu\'{e}bec, Canada, H3G 1M8}

\date{\today}
\address{}
\address{\mbox{}}
\address{\parbox{14cm}{\rm \mbox{}\mbox{}\mbox{}
The Coulomb drag between parallel, {\it ballistic} quantum wires is
studied theoretically  in the presence of a perpendicular magnetic field $B$.
The transresistance $R_D$ shows peaks as a function of the
Fermi level and splitting energy between the 1D subbands of
the wires. The sharpest peaks appear when the Fermi level
crosses the subband extrema so that the Fermi momenta are small.
Two other kinds of peaks appear when either {\it intra}- or
{\it inter}-subband transitions of electrons 
have maximum probability; the {\it intra}-subband transitions  
correspond to a small splitting energy.
$R_D$ depends on the field $B$ in a nonmonotonic fashion:
it decreases with $B$, as a result of the suppression of backscattering, and
increases sharply when the Fermi level  
approaches the subband bottoms and the suppression is outbalanced by
the increase of the Coulomb matrix elements and of the density of
states.}} 

\address{\mbox{}}
%\vskip 1truecm
\address{\parbox{14cm}{ \rm PACS: 73.20 Dx}}
\maketitle
\clearpage
\noindent

\section{ Introduction}

Experimentally \cite{1} and theoretically \cite{2} momentum transfer between
spatially separated electron layers or Coulomb drag has been studied mostly
between two-dimensional (2D) layers. Theoretically,
this drag has also been studied between very long one-dimensional (1D)
layers (quantum wires) in which the
wire length $L$ is much longer than the mean free path $l_i$
({\it diffussive} regime\cite{3}) and recently between 1D layers
of length $L\ll l_i$, in which the electron motion along the layer,
at low temperatures, is mostly ballistic \cite{4} \cite{5}
({\it ballistic} regime). Even when  most of the electrons
pass along the wires without collisions, a few of them
experience backscattering due to the interaction with the electrons
of the other wire and this modifies the time-averaged distribution
functions in such a way that the drag effect occurs. In both regimes
the drag response is found to be maximal when the subbands in the
two wires line up precisely. It is important that in the ballistic
regime the transresistance is determined only by the Coulomb
interaction between the electrons and such basic properties of the
layers as the number of occupied subbands, and does not include the
relaxation characteristics of the system such as scattering times.
Therefore, the ballistic regime provides a possibility to obtain
more direct information about the Coulomb interaction in 1D
electron systems.

Motivated by the  results of Refs.  \cite{4} and \cite{5} 
we undertook an extended theoretical study of the drag in the ballistic
regime, without tunneling between the wires, but in the presence of a
perpendicular magnetic field $B$. In  Secs. II and III, we generalize the
theory of Ref. 4 to
include the effects of intersubband transitions in electron-electron
collisions and  account for the influence of a magnetic field on
the Coulomb drag; a limited account of this influence, 
valid  when only the lowest subbands in the two wires are occupied, appeared
in Ref.  [5].  Concluding remarks follow in Sec. IV.

\section{General formalism}

We use a model of a four-terminal double-quantum-wire system,
as shown schematically in Fig. 1,
similar to the systems investigated in the "directional coupler"
problem \cite{6}. Two closely spaced quantum wires, numbered 1 and 2
and centered at $y_1$ and $y_2$,
are contacted independently to four leads at $x=0$ and $x=L$, where $L$
is the length of the wires. The leads have chemical potentials
$\mu_{1,2}(0)=\mu_{1,2}^+$, and $\mu_{1,2}(L)=\mu_{1,2}^-$. Applying the
bias $V=(\mu_{2}^+ -\mu_{2}^-)/e$ to the leads of wire 2 (drive wire) we
obtain the current $I$ flowing through it. This current induces
a voltage $V_D=(\mu_{1}^+ -\mu_{1}^-)/e$ in wire 1 (drag wire). This is
the typical setup for drag measurements \cite{1}. We assume that the
barrier between the wires is high enough to allow the neglect of tunneling.

Below we assume that the electrons in
each wire are parabolically confined by the potentials
$U_j=\varepsilon_{j}^0 +
m^{*}\Omega_j^2 (y-y_j)^2/2$, $j=1,2$. In the presence of a perpendicular
magnetic field $B$, introduced through the vector potential ${\bf A}=
(-By, 0,0)$, the normalized
wave functions are $\Psi_{jn k}(x,y)=e^{ikx} \chi_{j n k}(y)$,
 $\chi_{jn k}(y) = (\pi^{1/2} \ell_j 2^n n !)^{-1/2}
\ H_n ((y-Y_j)/\ell_j)) \exp (- (y-Y_j)^2/2 \ell_j^2)$, where
$n$ is the 1D subband number, $k$  the wave vector of electrons, and
$H_n (x)$ the Hermite polynomials. Neglecting spin splitting the
corresponding energy spectrum $\varepsilon_{jn}(k)$ reads  

\begin{equation}
\varepsilon_{j n k}=\varepsilon_{j}^0 + \hbar \omega_j (n+1/2) +
(\hbar^2/2 m_j) (k-y_j/\ell_c^2)^2.
%1
\end{equation}
Here $\omega_j^2=\omega_c^2+\Omega_j^2$, $\omega_c=eB/m^{*}$ is the
cyclotron
frequency, $m_j=m^{*} \omega_j^2/\Omega_j^2$, $\ell_c=(\hbar/m^{*}
\omega_c)^{1/2}$
is the magnetic length, $\ell_j^2=\hbar/m^{*} \omega_j$, and
$Y_j=[\Omega_j^2 y_j + \hbar \omega_c k/m^{*}]/\omega_j^2$ are the
$k$-dependent  centers of the oscillators.
The expressions for the kinetic energies $(\hbar^2/2 m_j)
(k-y_j/\ell_c^2)^2$ of the electrons can be simplified
by a gauge invariant transformation resulting in a
shift of $k$ by an arbitrary constant. Since we neglect tunneling, we
do not consider electron transitions between the wires 
and can make such shifts independently for each wire; this does not
affect the kinetic equations written below. Explicitly, we
will shift the wave vectors in the manner $k-y_1/\ell_c^2 \rightarrow k$ for
wire 1 and $k-y_2/\ell_c^2 \rightarrow k$ for wire 2.
Then the kinetic energies in Eq. (1) will read $\hbar^2 k^2/2 m_j$
and the  oscillator centers  $Y_j=y_j + (\hbar \omega_c/m^{*} \omega_j^2) k$.

If the distribution functions $f_{j k}(x)\equiv f_{j k}$
change over distances much longer than both the electronic
wavelength $\pi/k$ and the characteristic radius of the
interaction potential, we can write the Boltzman kinetic
equations as

\begin{eqnarray}
\nonumber
&&{\hbar k \over m_j} {\partial f_{j n k}(x) \over \partial x} =
-\frac{4 \pi}{\hbar} \sum_{j' k'q} \sum_{n n' n_1 n_1'}
\left| M^{jj'j'j}_{n_1 n_1' n' n} (k,k',q) \right|^2
\delta( \varepsilon_{j n k}+
\varepsilon_{j' n' k'} - \varepsilon_{j n_1, k-q}  -
\varepsilon_{j' n_1',k'+q} )\\*
\nonumber \\
%\nonumber \\
&&\times [f_{j n k} (1-f_{j n_1, k-q} ) f_{j' n' k'}
(1-f_{j' n_1', k'+q} )
%\nonumber \\
- f_{j n_1, k-q} (1-f_{j n k} ) f_{j' n_1', k'+q}  (1-f_{j'n' k'} ) ],
%2
\end{eqnarray}
where the collision integral accounts only for electron-electron
scattering. The Coulomb matrix elements $M^{jj'j'j}_{n_1 n_1' n' n}
(k,k',q)$ are given by\\

\begin{equation}
M^{jj'j'j}_{n_1 n_1' n' n} (k,k',q)= \frac{2 e^2}{\kappa}
\int dy  \int d y'
K_0(|q||y-y'|) \chi_{j n k}(y) \chi_{j'n' k'}(y') \chi_{j'n_1' k'+q}(y')
\chi_{j n_1 k-q}(y),
%3
\end{equation}
where $\kappa$ is the dielectric constant and $K_0$ the modified
Bessel function.

It is convenient to write separately the distribution functions for
the forward- and backward-moving electrons as $f^+_{j |k|}=f_{j k}|_{k>0}$
and $f^-_{j |k|}=f_{j k}|_{k<0}$, respectively. For these functions
the boundary conditions are given in the Landauer-Buttiker approach by
$f^+_{j n k}(0) = f(\varepsilon_{j n k} - \delta \mu_{j}^+)$ and
$f^-_{j n k}(L) = f(\varepsilon_{j n k} - \delta \mu_{j}^-)$, where
$\delta \mu_{j}^{\pm}=\mu_{j}^{\pm} - \mu$, $\mu$ is the equilibrium
chemical potential, $f(\varepsilon)=[e^{(\varepsilon-\mu)/k_B T} +1]^{-1}$,
and $T$ the temperature. For $j=1$ and $j=2$, Eq. (2) gives two
coupled kinetic equations whose solution allows us to express the unknown
potentials $\mu_{1}^-$ and $\mu_{1}^+$ through the fixed $\mu_{2}^-$
and $\mu_{2}^+$ values and thereby calculate the transresistance.

\section{ Results}

        If most of the electrons move through the wires balistically,
Eq. (2) can be solved by simple iterations. The zero-order
approximation gives $f^+_{j n k}(x) = f(\varepsilon_{j n k} - \delta \mu_{j}^+)$ and
$f^-_{j n k}(x) = f(\varepsilon_{j n k} - \delta \mu_{j}^-)$.
Substitution of these functions in the collision integral
gives a non-zero contribution for backscattering collisions between
the electrons of different wires. This is the main contribution
which will be considered in detail in the following. If more than
a single subband in a wire is occupied, the intersubband transitions
within one wire also contribute to the collision integral of Eq. (2)
(if only the lowest subband is occupied, the intralayer part of the
collision integral completely vanishes because of the relation
$q=k-k'$ following from the momentum and energy conservation rules).
Howewer, within the iterative approach used here, we can neglect
the influence of the intralayer collisions on the distribution
functions of the drive layer ($j=2$), since the transport regime
is nearly ballistic. Further, the intralayer collisions do not
modify considerably the distribution functions of the drag layer
($j=1$) because $\mu_{1}^+ -\mu_{1}^-$ is assumed to be much smaller
than $\mu_{2}^+ -\mu_{2}^-$, and the main effect on
$f^{\pm}_{1 n k}(x)$ results from the interlayer Coulomb interaction.
Considering only  contributions  linear  in $V$, we
substitute the equilibrium Fermi-Dirac functions $f^{\pm}_{1 n k}(x)
=f(\varepsilon_{1 n k})$ in the collision integral and obtain

\begin{eqnarray}
f^+_{1 n k}(x) = f (\varepsilon_{1 n k} - \delta \mu_{1}^+)
- eV (m_1/\hbar k) \lambda_n (k) x, \\
f^-_{1 n k}(x) = f (\varepsilon_{1 n k} - \delta \mu_{1}^-)
+ eV (m_1/\hbar k) \lambda_n (-k) (x-L),
%4,5
\end{eqnarray}
where the factor  
\begin{eqnarray}
\lambda_n (k)= \frac{ 4 \pi}{\hbar k_B T}
\sum_{n_1 n' n_1'} \sum_{k' q} \left\{ \left[
\left| M^{1221}_{n_1 n_1' n' n} (k,k',q) \right|^2
\delta( \varepsilon_{1 n k}+ \varepsilon_{2 n' k'}
- \varepsilon_{1 n_1, k-q}  - \varepsilon_{2 n_1',k'+q} )
\right. \right.
\nonumber \\
\left. \left. \times
f(\varepsilon_{1 n k}) [1-f(\varepsilon_{1 n_1 k-q})]
f(\varepsilon_{2 n' k'}) [1-f(\varepsilon_{2 n_1' k'+q}) ]
\right]_{k' >0, k'+q < 0} - \left[...\right]_{k' <0, k'+q > 0}
\right\}
%6
\end{eqnarray}
is determined by the Coulomb matrix elements and the equilibrium
distribution functions only. The current flowing in the drag wire
is given by

\begin{equation}
I_D= \frac{e}{\pi} \sum_n \int_{0}^{\infty} d k \ (\hbar k/m_1)
\ [f^+_{1 n k}(x)- f^-_{1 n k}(x)].
%7
\end{equation}
$I_D$ does not depend on $x$ due to the property $\sum_n
\int_{-\infty}^{\infty}\lambda_n (k) d k=0$, which follows from
detailed balance. Substituting Eqs. (4) and (5) into Eq. (7),
using the requirement $I_D=0$, and defining the transresistance
$R_D$ as $R_D=-V_D/I$ through the ballistic current $I=V/R_N$, where
$R_N=h/2e^2 N$ is the ballistic resistance of wire 2 and $N$
 the number   of occupied subbands, we finally obtain

\begin{eqnarray}
\nonumber
&&R_D=\frac{ \hbar L}{N N_D e^2 k_B T}
\sum_{n,n_1=0}^{N_D-1} \sum_{n',n_1'=0}^{N-1}
\int_{-\infty}^{0} d k' \int_{-k'}^{\infty} d q \int_{0}^{q} dk
\left| M^{1221}_{n_1 n_1' n' n} (k,k',q) \right|^2\\*
\nonumber
\ \\
&&\times\delta( \varepsilon_{1 n k} + \varepsilon_{2 n' k'}
- \varepsilon_{1 n_1, k-q}  - \varepsilon_{2 n_1',k'+q} )
%\nonumber \\
%\times
f(\varepsilon_{1 n k}) [1-f(\varepsilon_{1 n_1 k-q})]
f(\varepsilon_{2 n' k'}) [1-f(\varepsilon_{2 n_1' k'+q}) ]
%8
\end{eqnarray}
Here $N_D$ is the number of occupied subbands in the drag wire
(wire 1). Note that the introduction of both $N$ and $N_D$ assumes
that the theory is valid when the Fermi energy
%%$\varepsilon_{F}
$\mu - \varepsilon_j^0
- \hbar \omega_j (n+1/2)$ with respect to the highest occupied level
is larger than $k_B T$. This, of course, implies that the 1D subband
separations $\hbar \omega_1$ and $\hbar \omega_2$ are much larger
than $k_B T$ and is true at $T \sim 1$ K for   electrostatically
defined electron channels.

Below we consider the case of identical wires $\Omega_1
=\Omega_2=\Omega$, which entails $\omega_1=\omega_2=\omega$,
$\ell_1=\ell_2=\ell$, and $m_1=m_2$). To further evaluate
expression (8), it is convenient to detach
the contribution $R_D^{(1)}$ from $R_D$ that expresses the equality
$n+n'=n_1+n_1'$ for which the energy conservation law gives
$q=k-k'$. Then we have $R_D=R_D^{(1)}+R_D^{(2)}$ with\\

\begin{eqnarray}
\nonumber
R_D^{(1)}&=&\frac{m^{*3} k_B T L \omega^6}{N N_D \hbar^5 e^2 \Omega^6}
\sum_{n,n_1=0}^{N_D-1} \sum_{n',n_1'=0}^{N-1}
\frac{\delta_{n+n',n_1+n_1'}}{k_n k_{n'} (k_n+k_{n'})}
\ \frac{\Delta_{n,n_1^{'}}^2}  
{\sinh^2\Delta_{n,n_1^{'}}}\\* 
\nonumber
\\
&&\times\left| M^{1221}_{n_1 n_1' n' n} (k_n,-k_{n'},k_n+k_{n'}) \right|^2.
%9
\end{eqnarray}
Here $\Delta_{n,n_1^{'}}=[\Delta+\hbar \omega(n-n_1')]/2 k_B T$ and
$\Delta=\varepsilon_1^0-\varepsilon_2^0$ is the interwire
splitting energy between the lowest subbands. Further, $k_n= (\omega/\Omega)
[2m^* (\mu -\varepsilon_1^0 - \hbar \omega(n+1/2) )/ \hbar^2]^{1/2}$ and
$k_{n'}= (\omega/\Omega)[2m^* (\mu - \varepsilon_2^0 -
\hbar \omega(n'+1/2) )/ \hbar^2]^{1/2}$ are the Fermi wave numbers for the
%bands
states $1,n$ and $2,n'$, respectively. The part $R_D^{(2)}$,
corresponding to $n+n' \neq n_1+n_1'$, is obtained as\\

\begin{eqnarray}
\nonumber
R_D^{(2)}&=&\frac{m^{*} L \omega^2}{2 \hbar N N_D e^2 k_B T \Omega^2}
\sum_{n,n_1=0}^{N_D-1} \sum_{n',n_1'=0}^{N-1}
\int_{0}^{\infty} d k \int_{0}^{\infty} d k'
(1-\delta_{n+n',n_1+n_1'})/p(k,k')\\*
\nonumber
\ \\
\nonumber
&&\times\Theta (k k'+\omega^2 (n+n'-n_1-n_1')/\Omega^2 \ell^2)
%\nonumber \\
%\times
\left| M^{1221}_{n_1 n_1' n' n} (k,-k',q) \right|^2
f(\varepsilon_{1 n k}) f(\varepsilon_{2 n' k'})\\*
\nonumber
\ \\
&&\times[1-f((\varepsilon_{1 n k}+ \varepsilon_{2 n' k'})/2 + \Delta(k,k')/2
)]
[1-f((\varepsilon_{1 n k}+ \varepsilon_{2 n' k'})/2 - \Delta(k,k')/2 )]
%10
\end{eqnarray}
where $p(k,k')=[(k+k')^2/4+ \omega^2 (n+n'-n_1-n_1')/\Omega^2
\ell^2]^{1/2}$,
$q=(k+k')/2+p(k,k')$, and $\Delta(k,k')=\Delta+\hbar\omega(n_1-n_1') -
(\Omega/\omega)^2 \hbar^2 p(k,k') (k-k')/m^{*}$. The statistical factor
in Eq. (10) is small  unless $|\varepsilon_{1 n k}-\mu|$,
$|\varepsilon_{2 n' k'}-\mu|$, and $|\Delta(k,k')|$ are small
enough and  comparable to $k_B T$. This allows the integrals over
$k$ and $k'$ are to be carried out in narrow regions
around  $k_n$ and $k_{n'}$, respectively. We used the same property
to reduce the contribution $R_D^{(1)}$ to expression (9). Although the
requirement $|\Delta(k,k')| \sim k_B T$ imposes certain restrictions
on the values of $\mu$, $\Delta$, and $\omega$, the processes with $n+n'
\neq n_1+n_1'$ can give a considerable contribution to $R_D$, especially
for $|\Delta+\hbar \omega(n-n_1')| \gg k_B T$ and
$R_D^{(1)}$ small. We stress that the
previous theoretical work \cite{4} on
the Coulomb drag in the ballistic regime took into account only the
processes with $n=n_1$ and $n'=n_1'$, thus neglecting other processes 
completely from the beginning. The numerical calculations given below
demonstrate that this limitation is considerable in many cases.

If only the lowest subbands are occupied in each wire,
i.e., for $n=n_1=n'=n_1'=0$, the calculation of the transresistance is
considerably simplified. Only $R_D^{(1)}$ contributes to $R_D$
and Eq. (9) can be rewritten as
\begin{eqnarray}
\nonumber
R_D&=&\frac{2 e^2 m^{*3} \omega^6 L k_B T}{\pi \hbar^5\kappa^2 \Omega^6
k_1 k_{1'}(k_1+k_{1'}) }\ \ \frac{(\Delta/2k_B T)^2}{\sinh^2(\Delta/2k_B
T)}\\*
\nonumber \\
&&\times e^{-(\omega_c/\omega)^2 \ell^2 (k_1+k_{1'})^2 }
%\nonumber \\
\left[ \int_{-\infty}^{\infty} d u e^{-u^2/2} K_0 \left(
(k_1+k_{1'}) |d + \ell u | \right) \right]^2.
%11
\end{eqnarray}
Expression (11) is convenient for assessing the magnetic-field
dependence of the  transresistance $R_D$ . It directly demonstrates a
significant reduction [5] of the drag effect by the magnetic field $B$,
mostly due to the exponential factor. The decrease of $R_D$ starts as
$R_D(B)-R_D(0) \sim - B^2$ and becomes exponential with increasing $B$.
The physical reason for this decrease is  the suppression of
backscattering in electron-electron collisions as the oscillator centers
 for forward- and backward-moving electrons are
pulled apart by the magnetic field. The characteristic field $B_0$ for this 
suppression depends on the position of the Fermi level
$\varepsilon_F$ and is estimated as $B_0 \sim (m^{*}/e)
( m^{*}\Omega^3/\hbar)^{1/2}/(k_1+
k_{1'})$. If $\varepsilon_F$ is not far from the subband bottoms,
$B_0$ is big and the suppression is weak. If $\varepsilon_F$ is well
in between the 1D subbands, $B_0$ is estimated as 1 tesla for typical wire
parameters. However,   when $\varepsilon_F$, with the increase of $B$,
 approaches  the subband bottom, the opposite effect takes
place: the transresistance increases because the wave vectors $k_1$,
$k_{1'}$, and $q$ become progressively smaller and the suppression of
backscattering becomes less important than the increase of  
the Coulomb matrix element and of the density of states.
 
Below we present   numerical results for
the transresistance $R_D$, expressed in units of the fundamental
resistance $R_0=h/2e^2$, at $T=1.3$ K, $L=0.4$ $\mu$m,
$d=|y_1-y_2|=50$ nm, $\hbar \Omega=$4 meV, $m^*=0.067 m_0$,
 and $\kappa=13$. Figure 2 shows the dependence
of $R_D$ on the Fermi energy $\varepsilon_F$, defined as
$\varepsilon_F =
\mu -(\varepsilon_1^0+\varepsilon_2^0)/2$, calculated at $B=0$
and $B=1$ tesla.  The calculations were done assuming that
up to two subbands can be populated in each wire. As seen in part (a), for
$\Delta=0$  there are pronounced sharp peaks of $R_D$ when
 $\varepsilon_F$ crosses the bottoms of the first and second
subbands. The sharpness of the peaks is explained by a strong
enhancement of the Coulomb collision probability
when $k_n$, $k_{n'}$, and $q$ are small, cf. Eqs. (9) and (10).
 In Fig. 2 (b), for $\Delta=1$ meV,  one can
see three peaks; the middle one appears after $\varepsilon_F$
crosses the bottom of the second subband in wire 2. This peak exists
due to the processes with $n+n'\neq n_1+n_1'$. The third,
most prominent peak in Fig. 2 (b), appears after $\varepsilon_F$
 crosses the bottom of the second subband in wire 1, so that two
subbands in both wires are populated. The processes with
$n+n'\neq n_1+n_1'$ give the main contribution to this peak as well.

The application of the magnetic
field shifts the peaks to higher Fermi energies, due to the increased
confinement energy, and sharpens them  due to the suppression of
backscattering in the regions far from the subband edges. The
resulting decrease of the transresistance due to this
suppression is illustrated in Fig. 3
for one ($\varepsilon_F=3.5$ meV) and two ($\varepsilon_F=8$ meV) subbands
populated. These dependences  are non-monotonic: when, with
the increase of $B$, $\varepsilon_F$
approaches the first or second subband bottom,  $R_D$
starts to grow sharply. This dependence is basically the same for both
$\Delta=0$ and $\Delta=1$ meV.

Figure 4 shows the dependence of $R_D$ on the level splitting
$\Delta$ at several constant values of $\mu-\varepsilon_1^0=
\varepsilon_F -\Delta/2$. This means that the subband positions of
wire 1 remain constant with respect to the Fermi level
but those of wire 2 do not; this can be
experimentally achieved, for example, by changing the voltage of the
gate adjacent to wire 2 while keeping that adjacent to wire 1 at a
constant voltage.
The curves are plotted for one (a) or two (b) subbands populated in
wire 1 but for different $\mu$, far from (solid) and close to (dashed)
the upper populated subband edge. Both curves of Fig. 4 (a) show two
peaks: the sharp ones appear when the second subband of wire 2
becomes populated while the broad ones appear
when the second subband of wire 2 is aligned with the first one of wire 1,
at $\Delta=\hbar \omega \simeq$ 4.36 meV. A similar behavior is seen in
Fig. 4 (b). At large negative $\Delta$ only one subband is populated in
wire 2 while at $\Delta \simeq -1.5$ (solid) and $-0.5$ meV (dashed)
the second subband of wire 2 becomes populated as well. This transition
is reflected by strong and sharp peaks in $R_D$. Other strong peaks
appear at $\Delta=0$, when the subbands are aligned; note that on the
dashed curve such a peak merges with that at $\Delta \simeq -0.5$ meV
and is not resolved. The minor peaks in the regions of negative and
positive $\Delta$ exist due to the inter-subband transitions with
$(n,n_1,n',n_1')=(0,1,0,0)$ and $(1,0,0,0)$ and $(n,n_1,n',n_1')=
(0,1,1,1)$ and $(1,0,1,1)$, respectively.
The maxima of these peaks occur when $\Delta(k, k^{'}) \approx
\Delta(k_{n}, k_{n'})$, cf. Eq. (10), goes to zero. Thus, the
level-splitting dependence of $R_D$ shows a rich structure of peaks
indicating that both the {\it intra}- and {\it inter}-subband
transitions of electrons contribute to $R_D$.

All calculations described in this section were repeated for
different values of the interwire separation $d$. An increase
of $d$ considerably decreases the transresistance: $R_D$ drops
by more than one order of magnitude as $d$ varies from 40 to 60 nm,
mainly due to the dependence of the Bessel function on its argument.
However, all qualitative features presented above are preserved.

\section{Remarks and conclusions}

The treatment of the drag effect in the ballistic
transport regime demonstrates the salient properties of
electron-electron collisions in double-layer quasi-1D electron
systems. The reduced dimensionality dramatically decreases
the scattering probabilities at low temperatures due to the
restrictions imposed by the momentum and energy conservation
laws. As a result, the transresistance shows peaks as a function
of either the Fermi level position or the interlayer level
splitting energy. The peaks always appear when the Fermi level
crosses the bottom of a subband, so that a new subband $n$ is
involved in the scattering prosess; the Fermi wave number $k_n$
for this subband is small, the density of states is high, and this
results in a higher scattering probability. When subband $n$ is
aligned to another one, the conservation rules allow electron
transitions inside the subband $n$, the corresponding momentum
transfer $\hbar q \simeq 2 \hbar k_n$ is small, and the Coulomb
matrix element is large thus giving rise to an additional increase
of the peak. Next, the peaks appear when two subbands from different
layers are aligned; this favors transitions which conserve
the sum of the subband numbers, $n+n'=n_1+n_1'$, especially the
transitions between the electrons inside the aligned subbands, cf.
Eq. (9). Finally, peaks occur under special conditions, for
$\Delta(k_{n}, k_{n'}) \simeq 0$, cf. Eq. (10); this implies a maximum
probability for intersubband transitions with $n+n' \neq n_1+n_1'$.
Although the peaks associated with these transitions are usually weaker
than those under subband alignment, they give a considerable
contribution which cannot be neglected. The described rich structure
of the peaks is best seen in the level-splitting dependence of the
transresistance shown in Fig. 4.

A magnetic field $B$ applied perpendicular to the wire plane reduces
the overlap between the wave functions for forward- and backward-moving
electrons and thereby tends to suppress  electron-electron scattering.
This results in a decrease of the transresistance. In addition, the
application of $B$ modifies the quantization energies and leads to a
shift of the subbands with respect to the Fermi level. Since the
scattering probability increases when a subband edge comes close
to the Fermi level, the transresistance $R_D$ may increase with the
increase of $B$. Therefore, the dependence of $R_D$ on $B$ is basically
non-monotonic as shown in Fig. 3.

The results obtained here are valid  when the 1D electron gas 
in either wire is described as a normal Fermi liquid.
We used this model  because the wires are short, the transport is nearly
basllistic, and the properties of the 1D electrons are determined
by those of the 2D reservoirs they are injected from. The case
of the Coulomb drag between 1D electron systems 
described as Luttinger liquids has been studied in Ref. 7.
 
Concerning experimental results we are aware only of those
of Ref. \cite{8} where the transresistance $R_D$ was measured as
a function of side gate voltages controlling the confining
potentials of the parallel, submicron-long quantum wires, thus
allowing to change the positions of the 1D subbands with respect 
to the Fermi level, the interlayer subband splitting $\Delta$, the wire
widths $W_j$, and the interwire distance $d$. It was found that $R_D$
shows sharp peaks when the Fermi level crosses the bottom of a 1D
subband. When the gate adjacent to the drag wire was kept at a constant voltage,
 correponding to one populated subband in it,  the transresistance, as a
function of the voltage of the gate adjacent to the drive wire, showed
two peaks. The shape and position of these peaks permit us to identify
them with those  of Fig. 4 (a), since the  situation described by Fig. 4 (a)  
corresponds roughly to this  type of measurements.
These experimental results provide qualitative support for our theoretical
predictions. However, as no formal connection is made in our model
between the gate voltages and the parameters $\varepsilon_F$, $\Delta$,
$d$, and $W_j$, we cannot attempt a more detailed comparison. 
Such a connection requires a detailed knowledge of the
gate-induced modification of the double-wire confining potential
which could be obtained only by a self-consistent solution of the 
electrostatic problem for the three-gate structure investigated in Ref. 8.
We expect though that further experimental and theoretical work will test
sufficiently the drag in the ballistic regime and our results.

\acknowledgements

The work of PV was supported by the Canadian NSERC Grant No. OGP0121756.

%\clearpage

%\clearpage

\begin{figure}
\caption{Schematic diagram of a coupled quantum-wire device.}
\label{fig.1}
\
\caption{Dependence of the transresistance $R_D$ on the position
of the Fermi level for (a) aligned, $\Delta=0$, and (b) shifted,
$\Delta=1$ meV, levels in quantum wires. The dashed and solid curves
correspond to $B=0$ and $B=1$ tesla, respectively. The other parameters
are listed in the text.}
\label{fig.2}
\
\caption{Dependence of $R_D$ on the magnetic field $B$, with one
($\varepsilon_F=3.5$ meV) and two ($\varepsilon_F=8$ meV) populated
1D subbands, at $\Delta=0$ (a) and $\Delta=1$ meV (b).}
\label{fig.3}
\
\caption{Dependence of $R_D$ on the level splitting energy $\Delta$
at $B=1$ tesla when one (a) or two (b) populated subbands of wire 1 
remain constant with respect to the Fermi level, $\varepsilon_F-\Delta/2=
const$. (a): $\varepsilon_F-\Delta/2=$3.5 meV (dashed) and 4 meV (solid).
(b): $\varepsilon_F-\Delta/2=$7 meV (dashed) and 8 meV (solid).}
\label{fig.4}

\end{figure}

\end{document}